\documentclass[]{aa}  

\usepackage{graphicx}
\usepackage{txfonts}
\begin{document}

   \title{Expectations for Fast Radio Bursts in Neutron Star-Massive Star binaries}

   \author{K. M. Rajwade
          \inst{1}
          \and
         J. van den Eijnden\inst{2,3}
          }

   \institute{ASTRON, the Netherlands Institute for Radio Astronomy, Oude Hoogeveensedijk 4, 7991 PD Dwingeloo, The Netherlands\\
              \email{rajwade@astron.nl}
        \and
        Department of Physics, University of Warwick, Coventry CV4 7AL, UK  
         \and
             Astrophysics, Department of Physics, University of Oxford,
Keble road, Oxford, OX1 3RH, United Kingdom\\
             }

   \date{Received \today; accepted}


  \abstract
  {Recent observations of a small sample of repeating Fast Radio Bursts (FRBs) have revealed a periodicity in their bursting activity that may be suggestive of a binary origin for the modulation.}
  {We set out to explore the scenario where a subset of repeating FRBs originates in binary systems hosting a highly energetic neutron star and a massive companion star, akin to $\gamma$-ray binaries and young High-Mass X-ray Binaries.}
   {In this scenario, we infer observables, compare them with current observational constraints, and make predictions for future observations. Firstly, we specifically focus on the host galaxy properties and binary formation rates. Subsequently, we investigate the expected evolution of the rotation and dispersion measure in this scenario, the predicted birth-site offsets, and the origin of the persistent radio emission observed in a subset of these systems. }
   {The host galaxies for repeating FRBs favour the formation of neutron star-massive star binary systems but any conclusive evidence will require future discoveries and localizations of FRBs. The birth rate of high-mass X-ray binaries, used as a proxy for all considered binaries, significantly exceeds the estimated rate of FRBs, which can be explained if only a small subset of these systems produce FRBs.
   We show that under simple assumptions, we can reproduce the DM and RM evolution that is seen in a subset of repeating FRBs. We also discuss the possibility of detecting a persistent radio source associated with the FRB due to an intra-binary shock between companion star wind and either the pulsar wind or giant magnetar flares. 
   The observed long-term luminosity stability of the Persistent Radio Sources is most consistent with a giant flare-powered scenario However, this explanation is highly dependent on the magnetic field properties of the neutron star. }
   {With these explorations, we have aimed to provide a framework to discuss future FRB observations in the context of neutron star-massive star binary scenarios. In conjunction, we currently conclude that larger numbers of localisations and observations of repeaters will be necessary to conclusively suggest or rule out a connection between (repeating) FRBs and such binaries.}

   \keywords{fast radio bursts --
                stars:neutron --
                stars:magnetars
                X-rays:binaries
               }

   \maketitle
%

\section{Introduction}

The discovery of Fast Radio Bursts (FRBs) was a transformative event in the field of transient radio astronomy. FRBs are millisecond-duration radio flashes of cosmological nature. To date, more than a decade after their discovery, more than 600 FRBs have been detected and reported, of which 24 are known to repeat and 19 have associated host galaxies~\citep[see][for more details and references]{petroff2022}. Repeating FRBs are particularly vital in studying their still poorly-understood origin, as follow-up observations can enable localization, host galaxy identification, and searches for (multi-wavelength) persistent counterparts~\citep[see][and the references therein]{nicastro21}. This highlights the need to discovered larger numbers of FRBs, especially repeaters or non-repeaters with (future) instruments capable of measuring accurate positions for single bursts. 

This work is motivated by two recent major developments in FRB observations. Firstly, two repeating FRBs have been discovered to show periodicity in their activity cycle~\citep{rajwade2020, chimeperiodicity2020}: FRB 20121102A was observed to show activity windows with a $\sim 160$ day periodicity, while FRB 180916.J0158+65 showed such windows at a $\sim 16$ day period. These results are consistent with a binary system origin of the FRB, where orbital motion modulates the observed burst rate~\citep[see e.g.][although see \citealt{beniamini2020}, \citealt{zanazzi2020} and \citealt{Sridhar2021} for alternative, non-binary explanations]{ioka2020}. Secondly, the Galactic magnetar SGR J1935+2154 showed extremely bright ($\sim$MJy-ms) radio bursts with associated high-energy flaring ~\citep{bochenek2020,chime_galacticmagnetar}. In fact, several radio bursts spanning eight orders of magnitude where emitted by the magnetar, suggestive of a link between the known neutron star population and FRBs. 

These two independent findings $-$ activity periodicities that may possibly imply a binary nature of the FRB source and the magnetar connection $-$ warrant a detailed examination of systems combining both properties \citep[see also e.g.][]{tendulkar2021}: how do binary systems hosting a magnetar, or more generally a neutron star, compare to known FRB properties? The notion that many known Galactic neutron stars reside in binary systems further motivates this approach, as the same may hold for any neutron star energetic enough to power an FRB. 

In this paper, we explore the properties of binaries hosting a neutron star with those of known FRBs. We will also explore several predictions for FRB observables in this scenario: the positions of these systems in their host galaxies, the evolution of dispersion and rotation measure (DM and RM, respectively) with time, and the properties of their persistent (radio) counterparts. In this approach, we will be agnostic with respect to the specific underlying FRB model: we only assume that the neutron star is the system is energetic enough to power it. We start, in the next section, with a brief overview of types of galactic binaries hosting a neutron star, in order to motivate a more specific focus on neutron star $-$ massive star systems in the rest of this paper. This next section will also set out the approach and structure for the remainder of our exploration. 

\section{Setting the stage: a brief overview of neutron stars in binaries}

\begin{figure*}[!ht]
	\includegraphics[width=\textwidth]{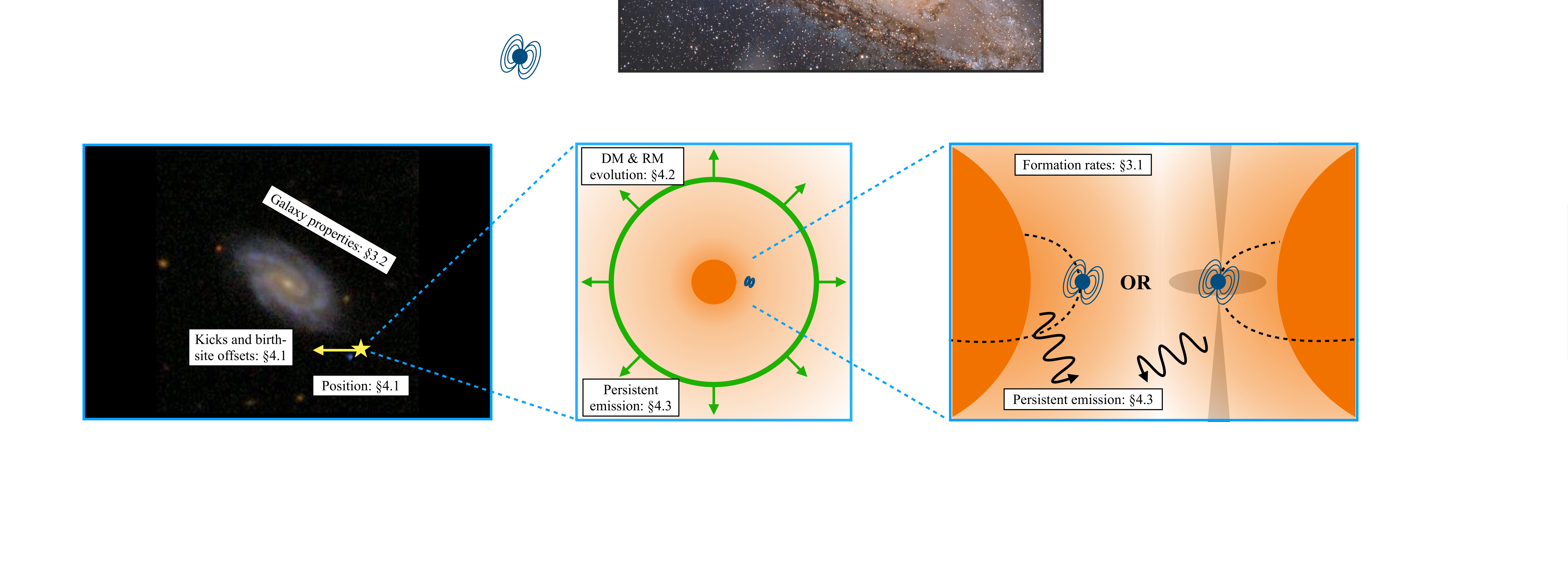}
    \caption{Schematic overview of the topics addressed in this paper and their structure in the text. We will discuss galaxy properties and formation rates of NMB-type FRB sources in Section \ref{sec:comparisons}. Subsequently, we discuss the expected kicks and offset, DM/RM evolution, and persistent radio sources, in Section \ref{sec:predictions}. The galaxy in the left panel is a \textit{gri} composite from SDSS, adapted from \citet{masters2019}. The middle panel shows a schematic depiction of an NMB surrounded by expanding material, while the right panel shows to configurations of a NMB, i.e. non-accretion (left) and accreting (right).}
    \label{fig:overview}
\end{figure*}

Galactic binaries hosting a neutron star and a non-degenerate companion star span a wide range of types, varying in a number of properties. Firstly, the two binary companions may interact via a range of processes, such as accretion \citep[i.e. X-ray binaries;][for low and high-mass X-ray binary reviews, respectively]{patruno2021,reig2011}, shocks between the pulsar wind and other components of the system \citep[e.g. stellar winds from the companion;][]{dubus2013}, or ablation of the non-degenerate star \citep[i.e. spider pulsars;][]{roberts2013}. In detached systems, on the other hand, such interactions do not take place. Across the wide zoo of neutron stars in binaries, basic neutron star properties differ greatly, as do those of the stellar companion: particularly, spin, magnetic field strength, and age of the neutron star, and mass, radius, and evolutionary status of the companion. Finally, the properties of the binary orbit itself, such as period and eccentricity, systematically differ between classes of binaries and systems of different age. 

Assuming that this Galactic population is representative of extragalactic populations, the two recent FRB discoveries introduced in the previous section rule out a significant subset of them as FRB sources. Restricting the systems to those consistent with a magnetar primary and with binary orbits in the range of $16$--$160$ days, we are left with young systems: specifically, those hosting a young, energetic neutron star and a massive donor star\footnote{We will discuss the recent localisation of an FRB in a globular cluster, i.e. and old population, in Section \ref{sec:discussion}.}. Here, the young neutron star matches with the magnetar constraint, while the binary orbits imply massive star companions\footnote{Note that this introduces an observational bias: non-interactive systems with a wide orbit and a low-mass companion may exist, but would not be observed as binaries. However, as we will discuss in Section \ref{sec:PRS}, a massive companion star may help to explain the presence of persistent radio sources at FRB positions.}.

In the Milky Way, two known classes of systems fit the description of binaries hosting a young neutron star and a massive star: high-mass X-ray binaries (HMXBs) and $\gamma$-ray binaries. The former are young systems, hosting strongly-magnetized ($B>10^{12}$ G) but slowly-spinning ($P>1$ sec) neutron stars \citep{staubert2019} accreting from a massive companion (OB supergiants or Be stars, typically) in an orbit with periods spanning from days to years \citep{reig2011}. The latter are similarly young systems, where the neutron star typically resides in an eccentric orbit around a massive star. While $\gamma$-ray binaries are sometimes classified as a subset of HMXBs, we distinguish them as no active accretion is thought to take place; instead their interaction takes the form of an orbitally-varying collision shock between the pulsar and stellar wind \citep{dubus2013}. The neutron star spin period, and derived dipolar magnetic field strength, are only known for three $\gamma$-ray binaries: PSR B1259-63 \citep{manchester1995}, LS I +61$^{\rm o}$ 303 \citep{weng2022}, and PSR J2032+4127 \citep{abdo2009}. The known $\gamma$-ray binaries also display orbital periods of similar order of magnitude as the two FRB activity periods. These $\gamma$-ray binaries furthermore show that, for orbital periods of the order of tens to hundreds of days, accretion does not necessarily take place, even in eccentric systems. Indeed, for the typical mass ratios implied in a neutron star $-$ massive star binary, most massive stars will not fill their Roche lobe \citep{underhill1979}. Therefore, depending on the combination of outflow and circumstellar disk properties of the massive star \citep[e.g.,][]{grundstrom2006} and the eccentricity \citep{martin2009}, some system will and others will not allow accretion to take place.

In the remainder of this paper, we will compare these neutron star $-$ massive star binaries (hereafter NMBs) to FRB sources. We will also assess the effect that the presence of the massive star may have on FRB observables. As stated, we are agnostic regarding the exact FRB mechanism — importantly, we do not assume it requires a binary companion. Therefore, the discussion in the remainder of this paper may only apply to a subset of FRBs. 

In Figure \ref{fig:overview}, we show a schematic overview of the structure of this paper. In the next section, we focus on comparisons: we will specifically compare existing constraints on formation rates of NMBs with the rates of FRBs (Section \ref{sec:formationrates}), and will compare host galaxy properties of localised FRBs with relations between NMBs and their host galaxy environment (Section \ref{sec:galaxies}). Subsequently, we will turn to predictions of a NMB scenario: at what position are they born and how do these positions evolve over time (Section \ref{sec:posvel}); how does the presence of the massive star affect the (time-evolution of the) rotation and dispersion measure (Section \ref{sec:dmrm}); and what persistent radio source properties may be expected (Section \ref{sec:PRS}). Finally, in Section \ref{sec:discussion} we discuss caveats that are at odds with an NMB scenario, after which we conclude in Section \ref{sec:conclusions}. 

\section{Comparisons: formation rates and host galaxy properties}
\label{sec:comparisons}

\subsection{Formation rates}
\label{sec:formationrates}

Let us first discuss constraints on formation rates of NMB systems and compare those with observed FRB rates. Generally, a stringent test of any progenitor model for FRBs is whether the birth rate of the proposed progenitor agrees with those observed rates. In the framework that we discuss here, the relevant observed FRB rate concerns their birth rate, inferred from observations, instead of the rate of bursts themselves, given that a subset of FRBs repeat. 

Firstly, we can approach this question by focusing on the FRB source itself. As relevant in the scenario investigated here, we consider the FRB to be produced by a strongly-magnetised neutron star (e.g. magnetars). For a Milky Way-like galaxy, we assume a the birth rate of magnetars of 2.2$\times$10$^{-3}$~yr$^{-1}$, based on estimates for magnetar birth rates in our Galaxy by several authors~\citep[see e.g.][and references therein]{gill2007}. While we know that isolated magnetars have the ability to produce bright radio bursts~\citep{bochenek2020,chime_galacticmagnetar}, we then have to correct this rate by the expected number of magnetars to be in a binary system with a high mass stellar companion. Very recently,~\cite{chrimes2022} have used Binary Population and Spectral Synthesis (BPASS) simulations to show that the number of magnetars that can be expected to be in binary systems in our Galaxy is anywhere between 5--10$\%$. Hence, the expected number of magnetars to reside in bound binary systems ranges between $1.1$--$2.2\times10^{-4}$~yr$^{-1}$. 

In order to estimate the fraction of bound systems with massive star companions, we estimate the fraction of bound systems with companion masses comparable to OB-type stars. Following the recent review of \citet{offner2022}, we can assume that, before the supernova creating the neutron star, the primary star follows an initial mass function with power $-2.3$, while the secondary's mass function is uniform between zero and the primary mass. In that case, under the prior information that the primary star was a massive star (e.g. producing a compact object), the probability of the secondary also being a massive star is of the order of tens of per cent. The total birth rate is corrected by this fraction to compute the fraction of bound magnetar systems with massive companions. Finally, to account for the beaming of FRBs away from the observer, we assume a beaming fraction similar to the Galactic radio pulsars of 10$\%$~\citep{faucher2006} to keep our simulations simplistic. Then, the expected birth rate of FRB progenitors that reside in NMB systems and visible from Earth can only be expected to be of the order of $\sim$10$^{-5}$ --10$^{-6}$~yr$^{-1}$. We highlight that the birth rate would be even lower if beaming is stronger for FRBs but the current estimates are highly unconstrained as the physical emission mechanism for FRBs is still unclear. This is consistent with the lack of a confirmed magnetar in a binary system with a high-mass binary companion in our Galaxy.

We can alternatively approach this question from the angle of HMXB formation rates. HMXBs are, or course, just one of the types of NMB. However, HMXB formation rates are best constrained from models and observations amongst interacting types. There are no constraints on non-interacting NMBs, making any estimates for these systems implausible. Hence, we first focus on the HMXB subset of NMBs for this comparison. Modelling of the Galactic HMXB population by \citet{iben1995} derived a formation rate of HMXBs of the order $2.6\times10^{-3}$~yr$^{-1}$, significantly larger than the above rate for magnetars with massive companions. Not all those HMXBs will necessarily host a neutron star primary. However, \citet{iben1995} note that the great majority of the modelled systems ($\gtrsim 84$\%) is expected to host a neutron star. Observationally, the Galactic population of HMXBs is indeed dominated by neutron star systems, of which a significant fraction hosts a Be-donor. Therefore, there is an $> 2$ orders of magnitude discrepancy between the formation rate of Galactic neutron star HMXBs, and magnetars with high-mass donors. This discrepancy is increased further by accounting for the population of $\gamma$-ray binaries amongst NMBs, which are not included in the above rate estimates: \citet{dubus2017} derive and expected number of $101^{+89}_{-52}$ Galactic $\gamma$-ray binaries, which, despite its large uncertainties, would constitute a significant fraction of the number of NMBs. 

The rate discrepancy between these two lines of reasoning may be reconciled by the extreme nature of the neutron star required to produce the FRB (as well as, possibly, the persistent radio source: see Section \ref{sec:PRS} with regards to the magnetic field and spin required in NMB scenarios for this persistent emission). 

A low formation rate of energetic, fast-spinning magnetars in binaries is suggested by the lack of their detection in our Galaxy. Therefore, most of the formed NMBs may simply host a neutron star that is either never energetic enough to be the FRB source, or only extremely briefly. In addition, orbital properties may reduce the discrepancy as well: for very short orbits, accretion may take place regularly, possibly suppressing neutron star - companion star interactions and magnetospheric FRB mechanisms; for wide-orbit systems, on the other hand, interactions between the two binary companions could become too weak to detect as a persistent source. Still, we can rougly estimate the upper limit on the fraction of NMBs that can host an FRB emitting magnetar to $\leq$1$\%$ based on the rate discrepancy. The fraction will be significantly lower if only a subset of magnetars are capable of producing FRBs. The current upper limit is consistent with the observed fraction of repeaters showing periodicity.

\subsection{Galaxy properties of HMXB systems and FRB hosts}
\label{sec:galaxies}

Building on the reasoning in the previous section, we can briefly turn our attention to the properties of the environment and host galaxy. The formation rate of NMBs, in relation to the properties of their environment, has been studied in particular for HMXBs, which we will therefore focus on below. It is known to be heavily dependent on star-formation rate (SFR) and metallicity \citep{dray2006} of the galaxy. For instance, the formation rate of HMXBs has been constrained in the Large and Small Magellanic Clouds (LMC and SMC), which both underwent a burst of star-formation in the last tens of Myr. For the SMC,~\cite{antoniou2010} infer a formation rate of 1 HMXB per $3\times10^{-3}$ $M_{\odot}$/yr of star formation, during a period of star formation taking place between 25 and 60 Myr ago. The formation rate in the LMC was found to be approximately 17 times lower by~\cite{antoniou2010}, at 1 system per $43.5\times10^{-3}$ $M_{\odot}$/yr of star formation, dominated by a star formation burst $6$ to $25$ Myr ago. The difference between these rates has been attributed to metallicity, where the lower metallicity of the SMC leads to an increased formation of HMXBs per unit of star-formation rate \citep[for comparison, note that the Milky Way, which has a higher metallicitiy, hosts a factor $\sim 50$ fewer HMXBs per units stellar mass;][]{majid2004}. At the same time, this lower metallicity also causes a higher fraction of systems hosting a Be-donor compared to OB supergiant donors, which may have effects on persistent radio source scenarios (e.g. Section \ref{sec:PRS}). Taking a complementary approach of considering X-ray luminosity functions, the relation between the total number of HMXB in a galaxy and its global SFR has also been established by comparisons of galaxies spanning a range of $\sim 50$ in SFR \citep{grimm2003,gilfanov2004}: higher SFRs are found to be correlated with larger numbers of HMXBs (which dominate the X-ray luminosity function at high X-ray luminosities). The results suggest that, in a NMB model for FRBs, these FRBs should occur preferentially in high SFR and low metallicity environments. 

\begin{figure}
	\includegraphics[width=\columnwidth]{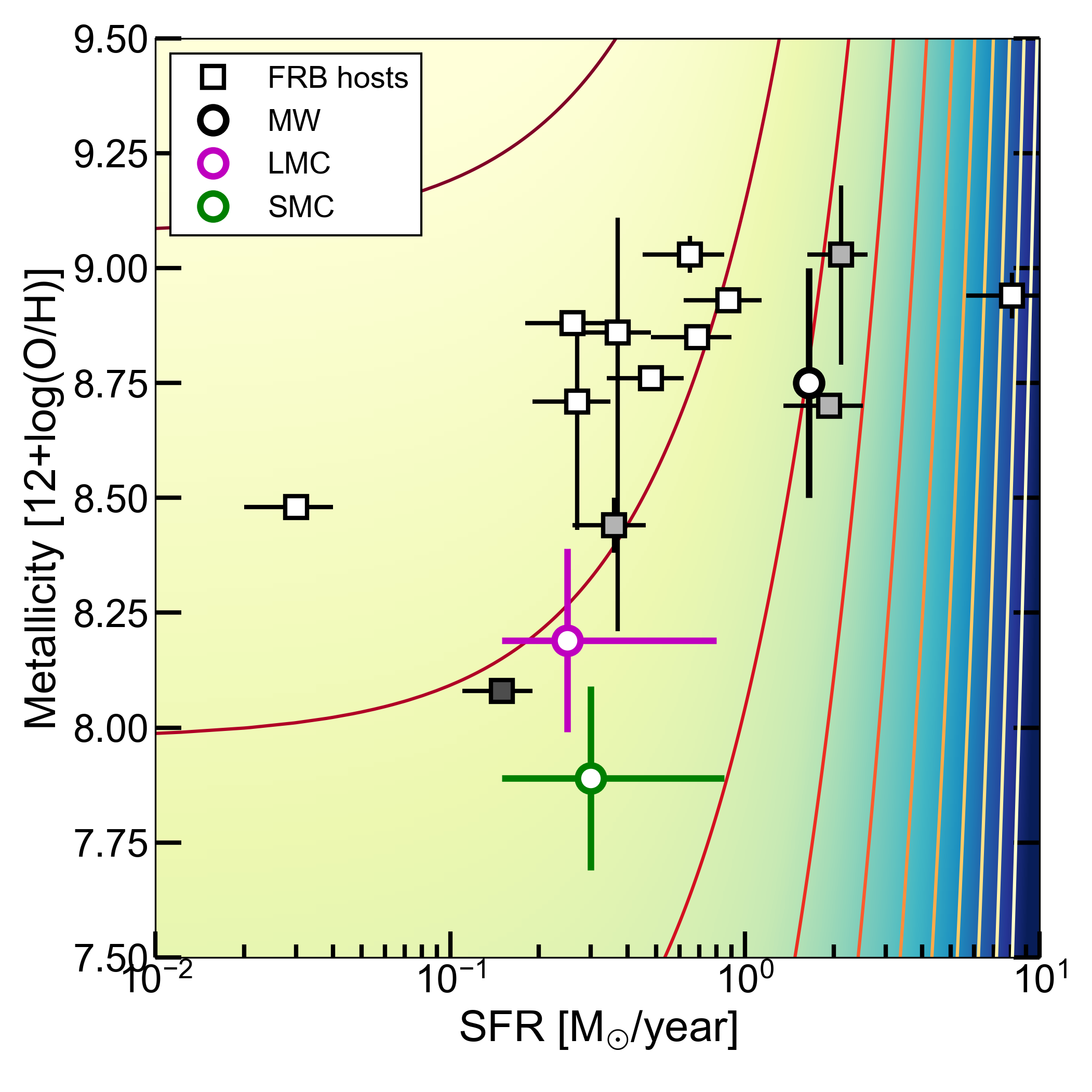}
    \caption{The SFR and metallicity of FRB hosts (non-repeaters in white, repeaters in grey filled points), alongside the Milky Way, LMC, and SMC. The colormap shows the dependence of the XLF on SFR and metallicity on a logarithmic scaling, from low (top left; light colors) to high (bottom right, dark colors). The lines indicate XLF contours increasing in steps of a factor ten, starting from an integrated X-ray luminosity of $10^{39}$ ergs~s$^{-1}$ for the top left contour. The dark grey-filled point is FRB~20121102A.}
    \label{fig:galaxies}
\end{figure}

More recent studies have shown that the relation between the number of HMXBs (or, more accurately, the X-ray luminosity function (XLF) of a galaxy), SFR, and metallicity does not change in a detectable manner between redshifts $z\sim 0.1$--$2$ \citep{douna2015,fornasini2020,lehmer2021} -- a range overlapping with the redshifts of identified host galaxies of FRBs. Therefore, we can compare host galaxy SFRs and metallicities of localized FRBs, when known, with the empirically determined relation between SFR, [O/H], and XLF: $\log L_x \propto \beta \text{[O/H]} \times \delta \log \rm SFR$, where $\beta = -0.91 \pm 0.17$ and $\delta = 1.06 \pm 0.08$ \citep{fornasini2020}. In Figure \ref{fig:galaxies}, we plot SFR versus 12+log(O/H) for the FRB host galaxies, while the background colormap traces the logarithm of the XLF, from low (light) to high (dark). The lines indicate contours in the XLF, increasing by a factor 10 between contours towards the (bottom) right. Finally, we include the Milky Way \citep[based on][]{balser2011}, LMC \citep{russell1992,antoniou2016}, and SMC \citep{russell1992,antoniou2010}, and show host galaxies of repeating FRBs in grey.

No clear trend is seen in the identified and characterised FRB hosts from Figure \ref{fig:galaxies}, as was similarly noted in analyses by \citet{heintz2020}, \citet{bhandari2020}, and \citet{lizhang2020} in comparing the properties of host galaxies with other possible FRB progenitor types. It is, however, interesting to note that the host galaxy of FRB~20121102A (dark grey square) lies very close to the LMC and SMC, galaxies that show a larger population of NMBs. As larger numbers of FRB hosts are identified and characterised in the future, a NMB model would predict those to lie preferentially towards the bottom right of this diagram. However, as discussed in the previous section, the extreme nature of the neutron star that may be required to produce the FRB may be the dominant and limiting factor in whether a galaxy host an FRB source. Another aspect to take into account is the time-scale on which the FRB-producing NMB evolves, compared to bursts of star formation: in the LMC and SMC, bursts of star formation tens of Myr ago are thought to be responsible for the current large populations of HMXB systems \citep{antoniou2010,antoniou2016}. Therefore, SFR estimates concurrent with FRB activity may not trace the presence of an FRB source, if the star formation activity lasted shorter than the lifetime of the massive binary before the first supernova (leading to the first of the FRB). However, those concerns may be more prominent for close-by systems such as the LMC/SMC, where burst of star-formation may be spatially resolved, instead of galaxies where we consider the averaged properties of the system. 

\section{Predictions: positions, DM \& RM, and persistent radio sources}
\label{sec:predictions}

Having compared the formation rates and host galaxy properties of (subtypes of) NMBs to localised FRB hosts, we will now turn to predictions in the NMB scenario. We will focus these predictions on three themes, where recent observations have revealed interesting and surprising FRB behaviour: the positions of FRBs in their galaxy and compared to potential birth sites; the time evolution of their DM and RM; and the presence of persistent (multi-wavelength) emission. The recent results in each of these three provide a benchmark for our predictive discussion, while future studies of larger samples can test the expectations for NMBs. 

\subsection{Position \& Offsets and birth sites}
\label{sec:posvel}

Through the combination of accurate localisation and deep imaging of their host galaxy, the potential birth site of three FRBs has been determined: FRB~20121102A \citep{marcote2017}, FRB~20180916B \citep{tendulkar2021}, and FRB~20190520B \citep{niu2021}. The possible offset between the FRB location and its birth place is consistent with a high peculiar motion of the FRB source, similar to Galactic compact objects (e.g. pulsar wind nebulae) and stars (e.g. runaway massive stars and hypervelocity stars). For an NMB system, such high velocities may be the result from the kick associated with the supernova that created the neutron star \citep[e.g.][]{blaauw1961}, or with many-body interactions between the massive star(s) and other stars in its original massive star cluster \citep{poveda1967}. 

Using the observed proper motion properties of Galactic high-mass X-ray binaries, we can assess the positional offset distribution expected for an NMB-type FRB source. For this purpose, we perform population synthesis simulations of the motion of NMBs within their host galaxy. As our observational input is based primarily on Galactic high-mass X-ray binaries, we assume host galaxy properties similar to the Milky Way. While isolated pulsars are typically born with large kick velocities due to the violent way in which they are created~\citep{hobbs2005}, this does not apply for high-mass X-ray binaries: \citet{bodaghee2021} found kick velocities for these systems in the range of 2--34~km~s$^{-1}$. These are the best kick constraints amongst the different types of NMBs and are the velocities projected over the sky plane~\citep[see][for more details]{faucher2006}. To account for the random direction of kick velocities in 3 dimensions, we draw the kick velocities from a zero-centred Gaussian distribution with a dispersion of 34~km~s$^{-1}$ over x, y and z-axis of the Cartesian co-ordinate system and then obtain the final velocity as a vector sum of the three components.

For each NMB, we then obtained its age by sampling from a uniform distribution between 0 and 10 Myr. In doing so, we assume that a NMB does not survive beyond 10 Myr, reflective of the time scale where the massive star is expected to end its life (we define that a binary becomes a NMB after the formation of a neutron star in the system). For the radial distribution of their original birth locations, we assume a distribution following the Galactic neutron star population~\citep{yusifov2004} $-$ effectively assuming that both pulsars and NMBs are correlated in similar fashion with star-forming region in our Galaxy. To simulate the time evolution of their positions, we then adopt an analytical model for the Galactic potential of the Milky Way: specifically, we adopt the the modification by~\citet{kuijken1989} of the~\citet{carlberg1987} fit of the Galactic gravitational potential. Thereby, we assume a combination of a disc-halo, bulge, and nucleus component, using the analytical forms and constants from~\citet{faucher2006}.

Then, we evolve the NMBs as follows:
\begin{itemize}
    \item For each NMB, we assign a birth velocity by drawing from the birth velocity distribution.
    \item We assign an age by drawing from the uniform age distribution of NMBs.
    \item We assign an initial birth position in the Galaxy by drawing from the radial and scale height distribution adopted from radio pulsars.
    \item For a given birth velocity and initial position of the NMB, we evolve it through the Galactic potential for a duration equal to its age to get the final position in the Galaxy.
\end{itemize}
For the purposes of these simulations and to get a statistically large sample of NMBs in the Galaxy, we simulate 200 systems using this method.

Finally, for each simulated NMB, we compute the offset from the birth site of the neutron star. To maximise the observed separation, we assume a face-on orientation of the galaxy with respect to the observer. For an original NMB position $X_0$, $Y_0$ at the time of the birth of the neutron star and final position $X_1$, $Y_1$ in a galactocentric Cartesian co-ordinate system, the offset with respect to the observer is simply $\sqrt{( X_1 - X_0)^2 + (Y_1 -Y_0)^{2}}$ parsec. 

\begin{figure}
	\includegraphics[width=\columnwidth]{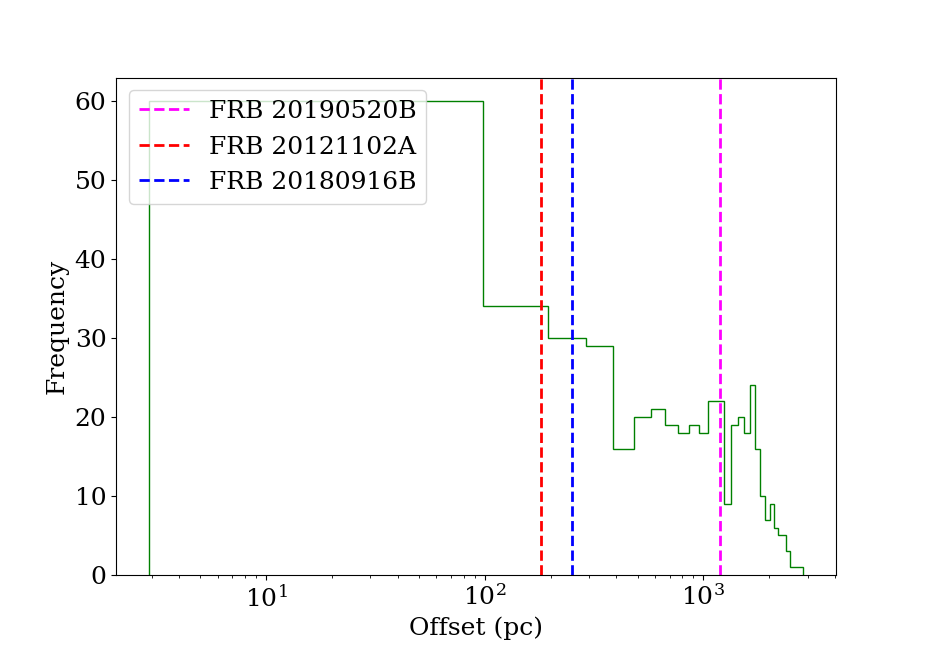}
    \caption{Histogram of projected offset distribution for the simulations of NMB-FRBs as described in section~\ref{sec:posvel}. The vertical lines shows the measured offsets measured for the three well-localized repeating FRBs showing that they are consistent with the expected distribution in an NMB scenario.}
    \label{fig:offset}
\end{figure}

To show the results of this simulation, we show a histogram of resulting offsets in Figure \ref{fig:offset}. The green curve indicates the histogram, while the three observed offsets, for FRB~20121102A, FRB~20180916B, and FRB~20190520B, are indicated by the vertical lines (converted from the observed angular offset between their assumed birth sites and their observed positions, using the host galaxy distance). We can see clearly how all three systems are consistent with the expected distribution for NMB type systems with typical Galactic kick velocities. We note that the age estimate we get for FRB 20121102A from the offsets is higher ($\sim$10$^{6}$ years) than what is postulated by other authors. One significant assumption in our estimate is that the offset measured for FRB 20121102A is from the centre of the star-forming region~\citep{bassa2017}, while the position of the FRB still lies within the half-light radius of the region. The inferred age will differ significantly depending the FRB source's original location within the entire region; hence, our simulation does not rule other estimates invalid. We note, also, that there is an obvious observational bias in the measurement of offsets: systems with a smaller offset will be more straightforwardly associated with a star-forming region. The three localised FRBs where an offset has been measured make up only a small subset of all localised sources and may be biased towards smaller offsets.

\subsection{DM \& RM evolution}
\label{sec:dmrm}

\begin{figure*}
	\includegraphics[width=0.5\textwidth]{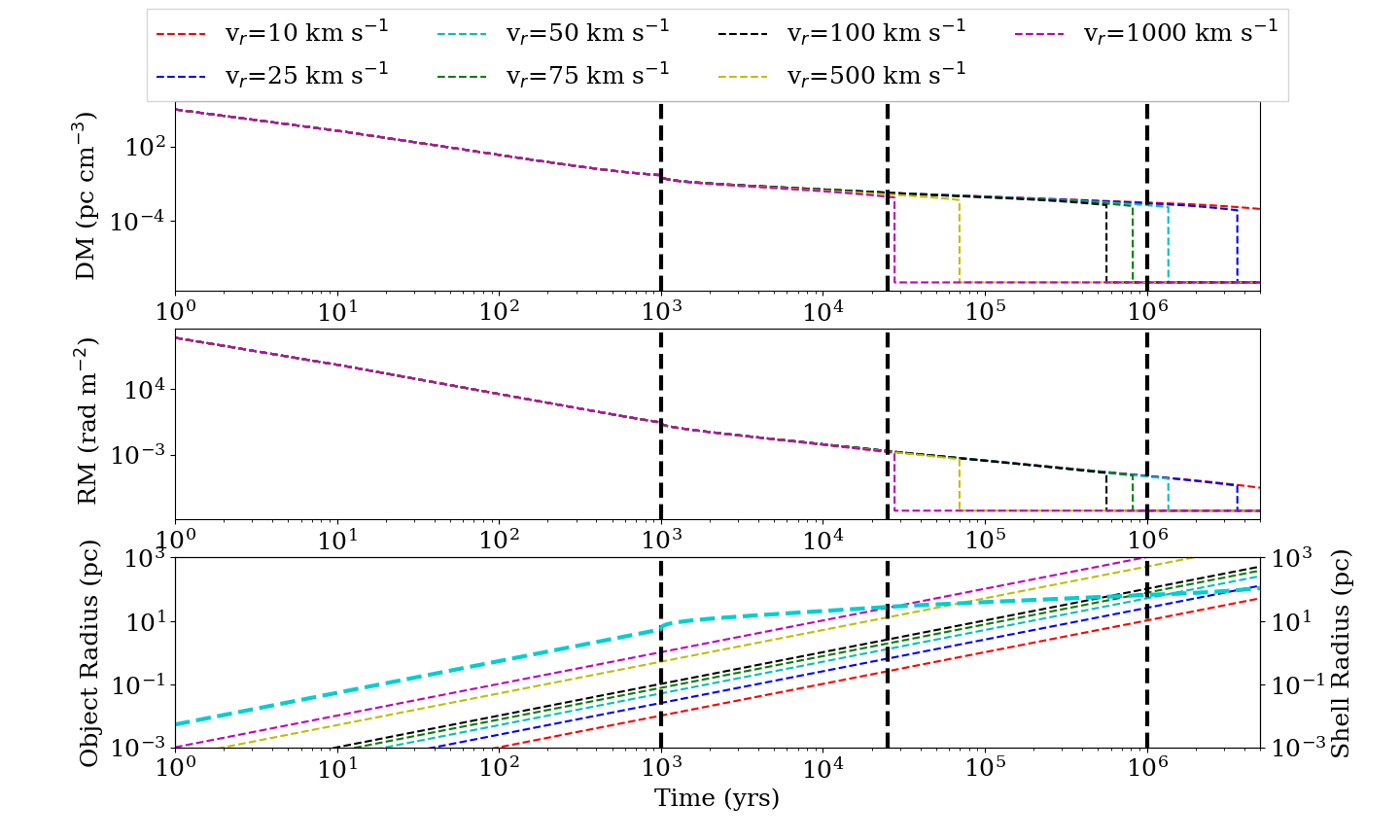}
	\includegraphics[width=0.5\textwidth]{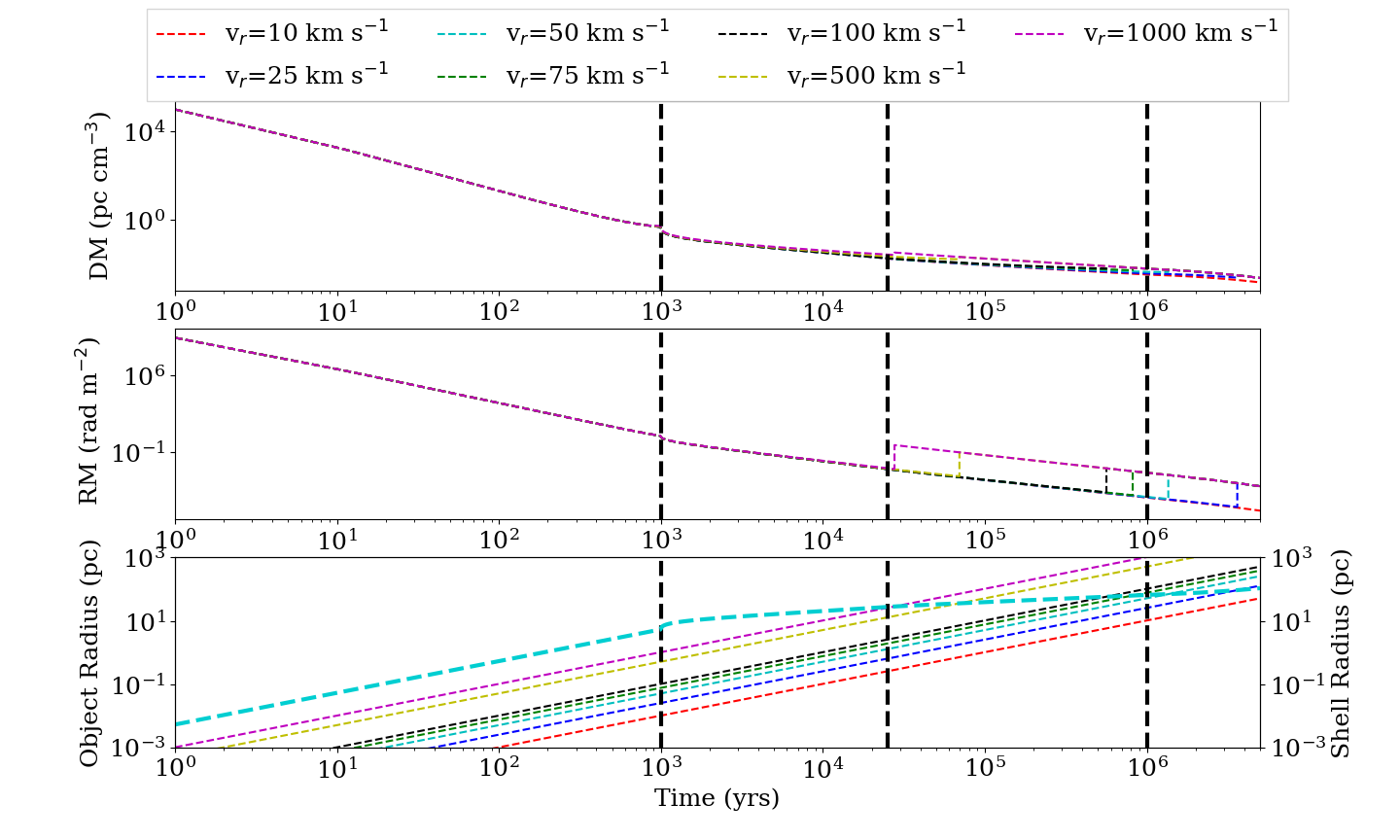}
    \caption{\textbf{Left Panel}: Three panels showing (from top to bottom) the DM, the RM and the radius of the SN shell as a function of time for a compact object moving toward the observer from the core of the shell with a certain radial velocity. The different colours denote the different radial velocities assumed for the compact object. The dashed vertical lines denote the times at which the ejecta velocity changes (corresponding to the evolutionary stage if the SN remnant). The dashed cyan line in the bottom panel shows the shell radius with time while the other lines correspond to the object radius (distance of the compact object from the centre of the SNR). \textbf{Right Panel}: Same plot as the left panel but for the compact object moving away from the observer.}
    \label{fig:dmrm}
\end{figure*}

Recent studies across time of several repeating FRBs have revealed variations in both the observed DM and RM, with significant variety in behaviour between sources. Monitored longest, FRB 20121102A can be characterised by the monotonic decrease in its RM over the span of 5--10 years, while its DM gradually increased at a rate of $\sim1$ pc cm$^{-3}$ year$^{-1}$~\citep{hilmarsson2021}. On the other hand, FRB~20190520B has shown erratic variations in RM characterised by sign changes and large variations in magnitude~\citep{anna-thomas2022}. Similarly, significant RM changes were observed in FRB~20201124A~\citep{wang2022}, while FRB~20180916B instead recently showed a secular rise in its RM~\citep{mckinven2022}. In an NMB-type scenario, both binary components can (indirectly) affect the RM and DM evolution with time: the remnant from the supernova that created the magnetar and the clumpy stellar wind from the massive companion star both supply ample material to interact with the FRB emission. While the former may naively be expected to cause a more gradual evolution over time scales of at least years, the latter may instead cause more erratic evolution on short time scales. In this section, we explore the effects that both structures may have on the FRB’s RM and DM. 

\subsubsection{The effects of a supernova remnant}
\label{sec:dmrm_snr}

To assess the effects of a surrounding supernova remnant (SNR) on the time evolution of RM and DM, we assume a magnetar created in a supernova explosion starts emitted FRBs after a short time compared to the evolutionary time scales of the remnant. As the SNR expands and, at the same time, the NMB may move within, through, and then out of the SNR due to a supernova kick, significant RM and DM evolution may be expected. Previous works by~\cite{piro2018} and~\cite{kundu2022} have performed analyses of how the SNR should contribute to the DM and RM of the FRB with time. While we refer to those work for their detailed treatments, we instead formulate a similar analysis that makes simplifying assumptions while capturing the salient features of the SNR evolution. Our main aim is to investigate the expected DM and RM evolution on a macroscopic and qualitative level, but note that the final and exact qualitative contributions of the SNR to the DM and RM will depend on the ionisation fraction, ejected mass, NMB kick velocity, and the interstellar medium (ISM) composition~\citep[see][for more details]{kundu2022}. 

Specifically, we make the following assumptions for a SNR evolving in time. Firstly, we assume that the SNR can be approximated a a spherically symmetric shell with a constant electron density within the shell surrounded by a thin layer of high-density swept-up material. For simplicity, we assume that the shell is mostly made up of Hydrogen and do not consider any heavier elements. Secondly, we assume that a reverse shock is created as soon as the ejecta hits the surrounding ISM material ($~$100~years since the event), after which is takes of the order $\sim 900$ years to propagate to the core of the remnant and ionise all the material within the shell~\citep{micelotta2016}. Until the time that reverse shocks reaches the core and completely ionizes the material within the shell, most of the material within the shell remains neutral (see Appendix A for more details about this assumption). Hence, we assume an ionization fraction of 10$\%$ until then. For this putative SNR, we use the measured values of various physical parameters based on the study by~\cite{micelotta2016} for Cas A, the best-studied Galactic SNR. We also evolve the ejecta velocity during different evolutionary stages (e.g. the free expansion stage, the adiabatic stage (Sedov-Taylor) and snow-plow phase), based on~\citep{Vink2020} and~\cite{wang2015} (see Appendix A for a detailed description of this calculation). Finally, we assumed the evolution of the magnetic field within the shell and at the shock front using the analytical treatment provided by~\cite{Vink2020}. Using this framework, we evolved the motion of a compact object created in this remnant, as well as the structure of the remnant itself, assuming a range of kick velocities. Assuming two scenarios, i.e. a compact object either moving towards or away from the observer, we then estimated the expected changes in DM and RM over time. 

In Figure~\ref{fig:dmrm}, we show the outcome of these calculations. In both columns, we show the DM (top), RM (middle), and SNR radius in comparison with NMB offset from the centre (bottom), all as function of time since the supernova. The left column denotes the calculations assuming the NMB moves towards the observer; the right column assumes the NMB moves away from the observer. Different colours of the dashed lines indicate different NMB velocities, while the black dashed lines indicate the times when the SNR is assumed to transition between evolutionary stages. 

Overall, we see that, when the NMB is moving toward the observer, the SNR expansion gradually causes a decay in both RM and DM over time. However, a clear drop in both, is predicted by the time the NMB leaves the SNR. For the case when the neutron star is moving away from the observer, the SNR expansion still causes a gradual decrease in RM and the DM. This can be attributed to the trade-off between the reduction in the electron density as the shell expands and the extra distance between the NS and the observer. In addition, once the NMB crosses the shell, more material intervenes the line of sight and an increase in both DM and RM is expected. This increase is, in a fractional sense, smaller than the decrease in the first scenario. We note how an arbitrary direction of the NMB between these two extremes will lead to an evolution in-between both scenarios, with no sudden change in DM/RM expected at the shell-crossing for motions perpendicular to the line of sight. 

This simplified calculation shows that one can replicate the DM and RM changes that are observed at least in some repeating FRBs. For example, the decrease in DM is fairly gradual during the last stage of evolution, which can be perceived as no detectable variability over a span of few years as seen for a few repeating FRBs~\citep[see][for e.g.]{pleunis2021}. Similarly, the RM contribution from the SNR decreases rapidly from being as high as $10^4$~rad~m$^{-2}$ to barely observable within the first 1000 years of the evolution. Rapid decreases in RM have indeed been observed for FRB~20121102A, while the RM of FRB~20180916B showed no variation over several years. Recently, however, its RM  increased monotonically~\citep{mckinven2022}, which may be attributed to an NMB moving through the dense, magnetic shell of the SNR in a direction away from the observer. If these FRBs originate from a NMB inside a SNR, the DM and RM evolution in them could be used as a probe of the age of the system. We finally note that the variations in DM and RM from these calculations do not require the presence of a binary companion, but merely the presence of a compact object formed in the supernova. Therefore, they could be testable on repeaters that do not show any signature of orbital motion. However, the presence of massive binary companion will systematically decrease the velocity of the FRB emitter, delaying the onset of sudden changes in DM and RM. 

\subsubsection{The effects of clumpy massive star winds}
\label{sec:dmrm_winds}

The effects of the SNR are however, clearly, unable to reproduce the more erratic RM behaviour observed in FRB~20190520B and FRB~20201124A: large changes in the magnitude of the RM and, especially, sign reversals. Such effects do now show up in our calculations and appear problematic for this interpretation. Therefore, one can instead consider the additional effects of the strong wind of the massive star in an NMB on the RM~\citep[see also e.g.][]{wang2022}: if this wind is clumpy, the line of sight may be crossed by a dense wind clump with a differently oriented magnetic field, leading to strong RM variations. 

\begin{figure}
\centering
	\includegraphics[width=0.49\textwidth]{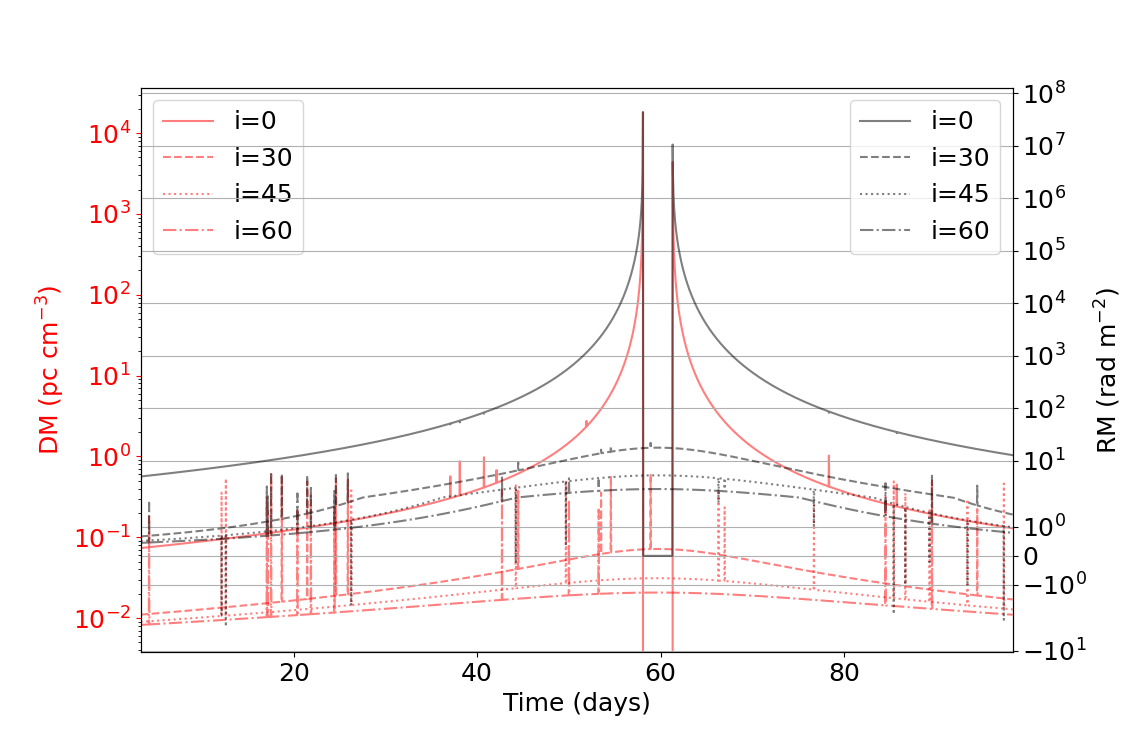}
    \caption{DM and RM contribution from the clumpy stellar wind as the FRB progenitor moves behind the companion. The sudden changes in the DM and RM values are due to the clumps in the stellar wind. The vertical lines correspond to the jumps in DM and RM due to the clumps in the stellar wind. The different curves are shown for different inclinations of the orbit (0 corresponds to edge-on orbit). The filling factor used for clumps for this simulation is 0.1.}
    \label{fig:dmrmbinary}
\end{figure}

The variations in DM and RM in an NMB system due to the presence of a stellar wind, may be a combination of secular changes dependent on the orbital phase and the orbital orientation with respect to the observer, as well as stochastic changes set by the wind’s clumpy structure. Such DM/RM variations are indeed seen in studies of radio pulsars in eclipsing binaries and other $\gamma$-ray binary systems~\citep{li2022}. To calculate these effects on a potential FRB emitter in orbit around a massive star, we assume that this magnetar is close to periastron in its orbit, while traversing behind the massive companion. From observational and modelling studies, it is known that massive stellar winds may reach filling factors as high as 0.1~\citep{puls2006}. To simulate such a clumpy wind, we first assume a baseline of a uniform wind where the electron density follows a radial profile of the type

\begin{equation}
    n_{r} = n_0~\left(\frac{R_0^{2}}{r^{2} - rR_0}\right) 
\end{equation}

where $n_{0}$ is the electron density close to the stellar surface, $R_{0}$ is the stellar radius, and $r$ is the distance from the star. We then add clumps stochastically with filling factors in the range of. 0.1--0.01 onto that steady wind at different time intervals; those interval are chosen in a random manner under the assumption that the generation of clumps in the wind is a stochastic process best characterised by a Poisson distribution. Then, the DM along the line of sight,
\begin{equation}
   {\rm DM_{\rm sw}} = \int_{R_{\rm NS}}^{R_{\rm out}} n_{r}\mathcal{S}(i, r, x)\, dr\,  + N\times n_{\rm clump}l,
\end{equation}
where $R_{\rm NS}$ is the distance of the neutron star from the companion, $R_{\rm out}$ is the outer limit up to which the stellar wind contributes to the DM ($\sim$100 stellar radii) and $\mathcal{S}(i, r, x)$ is the multiplicative factor that is required to convert the integral along our line of sight to a function of distance from the companion for an inclination angle ($i$)  and the projected distance between the NS and the companion ($x$). Here, $n_{\rm clump}$ is the electron density within the clump, $N$ is the number of clumps and $l$ is the path length of the clump along our line of sight.

We finally assume that magnetic field above the star evolves with distance as
\begin{equation}
    B_{r} = B_{0}~\left(\frac{R_{0}}{r}\right) \text{ ,}
\end{equation}
where $B_{0}$ is the magnetic field close to the surface of the star. The clumps in the wind can induce magnetic fields that can have different directions, leading to variations in the RM due to the field reversals in the clumps. To account for these reversals, we randomly assign a sign to the magnetic field in each of the the clumps. Similar to the DM, the RM contribution from the stellar wind,
\begin{equation}
   {\rm RM_{\rm sw}} = \int_{R_{\rm NS}}^{R_{\rm out}} n_{r}B_{r}\mathcal{S}(i, r, x)\, dr\, + \sum_{n=1}^{N} \pm 1 ~\times n_{\rm clump}B_{\rm clump}l,
\end{equation}
 where, $B_{\rm clump}$ is the magnetic field in the clump. The $\pm$~sign reflects the notion that the magnetic field component along our line of sight will have different signs for different clumps. We also ensure that the FRB is able to travel through this environment by computing the optical depth of the line-of-sight for the steady wind and the clumps. In order to do this, we compute the emission measure (EM) along with the DM for every line of sight \citep[e.g.,][]{wright1975}. Then, assuming an electron temperature of 10$^{4}$~K for the wind and the clumps, we compute the opacity at 1.4~GHz~\citep{draine2011} and find $\tau \ll$ 1, with values $>$ 1 only when the NS is very close to being eclipsed by the companion.
Assuming a velocity of the compact object along its orbit (i.e. compared to the stellar wind) to be 50~km s$^{-1}$ for typical NMBs with orbital period of few years~\citep{frank2002}, we then simulate the DM and RM variations one measures as the object moves towards and away from the periastron. The nominal values for $B_{0}$ and $n_{0}$ and other parameters of the massive star are taken from Table 1 of~\cite{zhao2022}. Figure~\ref{fig:dmrmbinary} shows the secular changes combined with sudden rises and falls in the DM and RM values. If the inclination of the orbit is high, one can still see erratic change in RM without any discernible changes in DM similar to similar to FRB~20190520B~\citep{wang2022,anna-thomas2022}. The magnitude of change in RM and DM in the clumps is dependent on the filling factor of the clumps and will be larger for denser clumps. We also note that the average direction of the magnetic field can also flip if the massive star has a circum-stellar disc. The disc has a strong toroidal magnetic field that would manifest as a sign change in the RM as the neutron star passes through the different sides of the disc~\citep{wang2022}.

\subsection{Persistent emission}
\label{sec:PRS}

In addition to detections of RM and DM variability, recent observations have also revealed persistent radio sources (PRSs) associated with two repeating FRBs. With the exception of the Galactic source SGR 1935+2154, no persistent (or burst) counterparts have been identified at other wavelengths as of yet. In this section, we will compare known persistent radio emission mechanisms of Galactic NMBs to the observed FRB PRSs and discuss a possible alternative origin in the NMB scenario. Similar to Section \ref{sec:dmrm}, we will also briefly mention the role that the SNR, expected around a young NMB, may play. However, let us first summarise the relevant known observational properties of PRSs. 

The PRS for FRB~20121102A was first detected, with a spectrum consistent with optically thin synchrotron emission and a specific luminosity of $\sim 2\times10^{29}$ ergs~s$^{-1}$~Hz$^{-1}$~\citep{marcote2017}. More recent observations of this PRS by~\citet{plavin2022} reveal a compact source down to milli-arcsecond scales, implying an intrinsic source size of < 1~pc. More recently, \citet{niu2021} reported the identification of a PRS associated with FRB~20190520B, at a specific luminosity of $3\times10^{29}$ ergs~s$^{-1}$~Hz$^{-1}$. Not all searches reveal a PRS, however: \citet{ravi2022} and~\citet{piro2021} report the detection of radio emission coincident with the position of FRB20201124A. However, both studies conclude that this emission likely originates from a star-forming region. \citet{ravi2022} therefore place an upper limit on the PRS luminosity of $3\times10^{28}$ ergs~s$^{-1}$~Hz$^{-1}$. In this context, FRB20180916B is similarly noteworthy: no PRS is known, but~\citet{tendulkar2021} found the FRB to be offset by $\sim 250$ parsec from a star-forming region where it may have originated between $800$ kyr and $7$ Myr ago.

The PRS of FRB~20121102A is best studied, especially across time due to its earliest discovery: upon its discovery,~\cite{chatterjee2017} report radio monitoring at $3$ GHz across approximately two separate months, which are spaced out by approximately three months. This monitoring campaign does not show evidence for a systematic increase or decrease of the PRS radio luminosity, but does show apparently less-structured variability around its mean luminosity. \citet{plavin2022} recently confirmed the absence of a luminosity decay on time scales of a year, which challenges several models for the nature of the PRS; particularly those invoking a transient radio counterpart powered by a single instance of energy injection (this possibility might still be feasible as we explain later). In the remainder of this section will take these basic observables — luminosities and (lack of) variability — to discuss the possibility of a PRS in the NMB scenario.

\subsubsection{Basic considerations on the emission origin}
\label{sec:prs_basicintro}

Here, let us initially consider a number of possible emission origins before considering two options in more depth for the NMB scenario. Firstly, the transient radio emission from either a supernova or SNR, creating the FRB source, has been considered by various authors to model various aspects of the PRSs. Such scenarios are more general that the NMB case consider in this paper, but the neutron star in the NMB implies a supernova has taken place. Observationally, it is known that radio-detected SNe typically rise in radio luminosity on time scales of days to hundreds of days, before subsequently decaying again. The radio-brightest examples peak at luminosities just above $10^{28}$ ergs~s$^{-1}$~Hz$^{-1}$, roughly an order of magnitude below the two confirmed PRSs. 

Recently,~\cite{eftekhari2019} reported the late-time radio brightening (after $\sim 7.5$ years) of the superluminous SN (SLSN) PTF10hgi, located at a redshift of $z=0.098$. Its peak radio luminosity is similar to the radio-brightest other SNe, around $10^{28}$ ergs~s$^{-1}$~Hz$^{-1}$, again below that of PRSs. In addition,~\citet{hatsukade2021} report that in new observations, PTF10hgi has decreased in radio luminosity, suggesting its radio peak has occurred. Despite a number of searches across a range of time scales since the explosion, no other radio detections of SLSNe have been obtained~\citep{mondal2020, hatsukade2021,eftekhari2021} -- except for a handful of radio detections that are consistent with host galaxy radio emission. Several searches for FRBs from the positions of SLSNe have also failed to detect any~\citep{law2019, hilmarsson2021}. While the transient nature of radio emission from SLSNe may also be challenging to reconcile with the stability of the FRB~20121102A PRS, a larger sample of radio-detected SLSNe would help to further assess any possible connection.

Turning towards NMB-specific scenarios, a radio counterpart similar to those of Galactic HMXBs can be ruled out easily. Those accretion-driven radio sources (i.e. powered likely through the launch of relativistic jets from their accretion flow) have been shown to be order of magnitude radio fainter than PRSs \citep{vdE2018,vdE2021}. Even the handful of known, extreme, young X-ray binaries (such as Circinus X-1 or SS 433) do not reach the required luminosities in a sustained fashion; furthermore, the nature of their compact object remains unclear but is highly unlikely to be similar to the magnetar invoked for the NMB scenario. Finally, \citet{sridhar2022} argue how radio nebulae created by the feedback of such extreme X-ray binaries would require mass accretion rate far exceeding those seens in the extreme Galactic X-ray binaries.

In our Galaxy, $\gamma$-ray binary systems are known to be radio-variable and bright systems. Their radio emission is not thought to be accretion driven, but instead shock driven: likely through shocks between the pulsar wind and stellar wind or decretion disk of their massive Be companion star \citep[see][for an extensive review]{dubus2013}. This motivates us to further explore the role such shocks may play in explaining PRSs, particularly for systems hosting young and extremely energetic magnetars — transporting the required energy either via their pulsar winds or giant magnetar flares. We note briefly that recently, scenarios involving a pulsar wind nebula (PWN, also referred to as a plerion) have been invoked as the explanation for the PRS \citep[e.g.][]{murase2016,kashiyama2017,chen2022}. While we do not explore this here in detail, as it does not require a binary companion, it similarly poses that pulsar spin down energy in converted to radio emission in shocks. However, those shocks then take place in the ISM instead, and could therefore be seen as a complementary option to our further discussion.

\subsubsection{Persistent luminosities of shocks powered by spin down or giant flares}
\label{sec:prs_lr}

Here, we will first consider the energetics and time-averaged power budget of particle-accelerating intra-binary shocks, powered either by magnetar spin down or giant flares. Starting with the former, we can consider the shocks taking place between pulsar wind and stellar wind/decretion diks, emitting non-thermally radio to very-high energy $\gamma$-rays \citep{dubus2013}. The emission is these systems is therefore thought to be powered, fundamentally, by the spin down energy of the neutron star. The physical sizes of the resolved radio emitting regions in $\gamma$-ray binaries are of the order $\sim 120$ AU~\citep{moldon2011}, significantly smaller than size upper limits of FRB PRSs. Their radio spectra are optically thin synchrotron spectra, consistent in spectral index with constraints in PRSs. An obvious discrepancy arises, however, when comparing the typical luminosities of both systems: $\gamma$-ray binaries are, at their brightest, several order of magnitude radio fainter than detected PRSs. Galactic $\gamma$-ray binaries are likely significantly older sources than FRB PRSs: PSR B1259-63, for instance, has a spin-down age of $330$ kyr and an inferred bipolar magnetic field strengths of $B \approx 3.3\times10^{11}$ G. Therefore, we can explore a scenario whether a very young, rapidly spinning magnetar in a $\gamma$-ray-binary-like configuration may power an FRB PRS.

Initially ignoring time variability (see Section \ref{sec:timeevo_lr}), the typical radio luminosity of PSR B1259-63 is approximately $6\times10^{29}$ erg/s at $2$ GHz, or $3\times10^{20}$ ergs~s$^{-1}$~Hz$^{-1}$ \citep{dubus2013}. Of all $\gamma$-ray binaries, its spin down energy is constrained best, at $\sim 8\times10^{35}$ erg/s \citep{manchester1995}\footnote{Recently, a spin measurement was reported for the $\gamma$-ray binary LS I +61$^{\rm o}$ 303 \citep{weng2022}. However, no spin derivative has been measured yet.}. For comparison, the PRS of FRB~20121102A instead has a specific radio luminosity approximately $9$ orders of magnitude higher, and would therefore require a spin-down energy of the order $\sim 8\times10^{44}$ erg/s if we assume the same efficiency between spin down power and radio luminosity of the inter-binary shock. Such a spin-down energy upper limit can only be feasible if a millisecond magnetar is formed in the system: for a magnetic field of $B=10^{14}$ G, it would require a spin frequency $\nu > 211$ Hz. On the other hand, assuming the pulsar spin does not exceed $\sim 2\times10^3$ Hz, the magnetic field should exceed at least $\sim 10^{12}$ G. Such magnetars have been proposed as progenitors of repeating FRBs~\citep[for e.g. see][]{margalit2018}. While this scenario is extreme, we note that the $156.9$-day activity cycle in FRB~20121102A is significantly shorter than the $1236$ day orbital period in PSR B1259-63. As the shock power decreases with stand-off distance from the pulsar, $\sim 8\times10^{44}$ erg/s could be considered an upper limit in this interpretation; the actual value could feasibly be (more than) an order of magnitude lower. 

Since $\gamma$-ray binary shocks emit across the entire electromagnetic spectrum, we should also consider whether detectable high energy emission may be expected in this scenario. In the X-ray band, we can again scale up Galactic $\gamma$-ray binaries for this comparison. Scaling the typical X-ray flux of PRS B1259-63 to the distance of FRB~20121102A, while also including the $\sim 9$ orders of magnitude higher normalization due to the extreme required spin down energy, one may expect expect a $\sim 10^{-14}$ erg/s/cm$^2$ X-ray flux for FRB~20121102A in this $\gamma$-ray binary scenario. This value slightly exceeds the X-ray upper limit from~\citep{scholz2017}, although this can be reconciled if the interstellar absorption $\sim 3$ times higher then assumed when deriving that upper limit (which remains consistent with the $N_H$ -- DM relation from \citealt{he2013}). Furthermore, the X-ray to radio luminosity ratio of PSR B1259-63 is significantly higher than for other $\gamma$-ray binaries. At their lower ratios, the X-ray flux of FRB~20121102A would indeed fall far below the observed limit. The latter effect may strongly depend on orbital separation, which affects the X-ray/radio ratio of the intra-binary shock. However, deep X-ray observations with future, low-background X-ray observatories could provide better X-ray tests of this shock scenario.  

At current $\gamma$-ray sensitivities and, especially, angular resolutions, the detection of an extragalactic persistent $\gamma$-ray counterpart is not expected. While the emission of Galactic $\gamma$-ray binaries peaks in the ten--hundreds of MeV band, it is limited by the pulsar spin down energy, while this $\gamma$-ray flux varies along the binary orbit. In the above scenario, therefore, the expected time-averaged $\gamma$-ray flux remains undetectable in e.g. \textit{Fermi}/LAT surveys. In addition, the relatively low angular resolution of $\gamma$-ray instruments challenges the confident identification of a persistent counterpart; indeed, as-of-yet, no constraining $\gamma$-ray limits on persistent emission have been reported.

A possible effect in this intra-binary shock scenario may be ablation effects of the massive companion star. Ablation is known to take place in spider pulsars, where a very-low-mass companion of a millisecond pulsar is slowly stripped of its outer layers by the energetic pulsar wind. A millisecond magnetar with an extreme spin-down energy may similarly ablate and strip a massive donor star. Following the formalism proposed by \citet{ginzburg2020} and \citet{ginzburg2021}, one can derive that such a spin-down power greatly reduces the time scales of effective ablation. While important differences exist between spider pulsars and NMBs -- the massive star is expected to reside more deeply within its Roche lobe and launches its own wind that shocks and balances the pulsar wind -- these ablation effects pose a challenge for spin-down-powered PRS scenarios.

Instead of pulsar spin down, the intra-binary shock may alternatively be magnetically powered, via the equivalent of a giant magnetar flare. The large amount of energy released in such a giant flare could lead to a persistent radio source, as the highly-energetic accelerated particles encounter the massive star's stellar wind. In an intra-binary shock between this wind of energetic particles and the stellar wind, synchrotron emission may arise as particles are trapped by and then gyrate in the wind's magnetic field. For instance, the Galactic magnetar SGR 1806-20 emitted a giant flare that injected $\sim$10$^{46}$~ergs of energy in its environment~\citep{palmer2005}. A radio afterglow was also detected after this event that showed rebrightening over the timescale of days~\citep{gelfand2004}. While the radio luminosity of the afterglow is again orders of magnitude lower than typical PRSs, one can explore a scenario where a PRS is powered for a longer period of time from a giant flare from a magnetar with a much larger magnetic field. While the radio afterglow in SGR 1806-20 was likely caused by shocks in swept-up ambient material, the presence of a massive star and its wind also provide a denser medium for the formation of such shocks. 

\begin{figure}
	\includegraphics[width=\columnwidth]{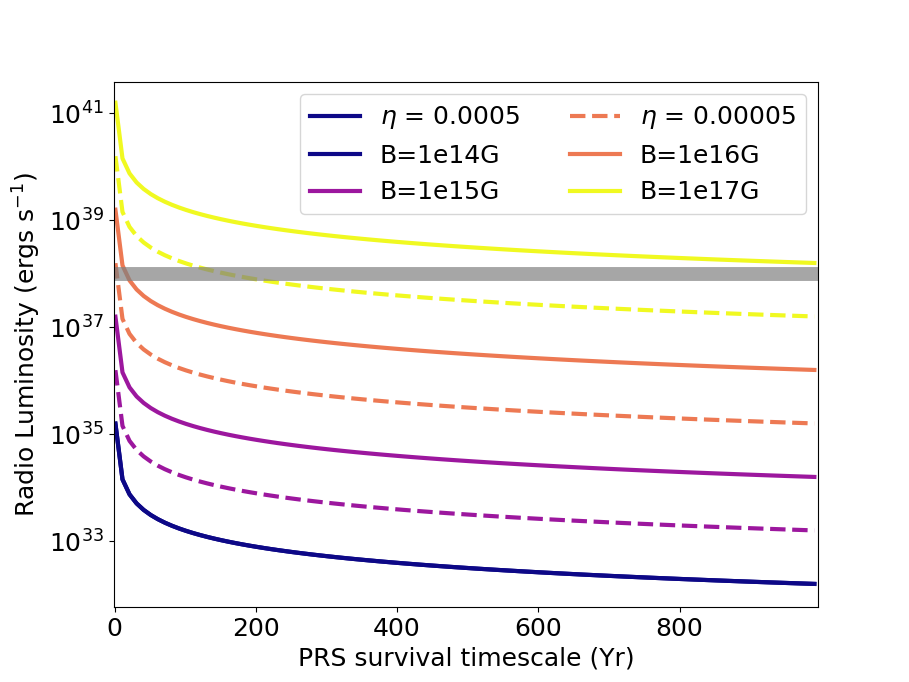}
    \caption{Radio luminosity of the PRS as a function of the lifetime of the PRS assuming no time evolution of the radio luminosity, for the case where a magnetar giant flare powers a PRS from the
    interaction with the stellar wind of the companion. Solid and dashed
    lines correspond to different efficiency of conversion to radio
    emission ($0.05$ and $0.005\%$, respectively) and different sets of
    lines correspond to different surface magnetic fields of the neutron
    star as shown in the legend. The grey horizontal region indicates
    the observed PRS luminosity of FRB~20121102A taken
    from \citet{chen2022}.}
    \label{fig:magPRS}
\end{figure}

To consider more quantitatively whether the PRS can be powered by magnetar flares, we here make the assumption that the total energy emitted during a giant flare is proportional to the square of the magnetic field. The aforementioned Galactic magnetar SGR 1806-20, that emitted the 10$^{46}$~erg giant flare, has an estimated magnetic field of $\sim$10$^{14}$~G. We then assume that flare-induced shock is powered by the interaction between the magnetar wind emitted during the flare and the stellar wind of the massive companion. The particles in the resulting shock are not expected to diffuse out quickly, as they are confined by the stellar wind’s magnetic field. As a consistency check, we computed the Larmor radius of the electrons for typical values of the Lorentz factor (10--1000). The Larmor radius is $\sim$ $3.3\times\frac{\gamma~mc^{2}~v_{\rm sp}}{qB}\,$m where $mc^{2}$ is the rest mass in the units of GeV, $v_{\rm sp}$ is the velocity perpendicular to the magnetic field, $q$ is the charge is units of the elementary change and $B$ is the magnetic field in Tesla. For electrons with GeV energies, the larmor radius is 3.3$\times$10$^{5}$~$v_{\rm sp}$~km. For large speeds (0.1--0.5c), we obtain Larmor radius of $\sim$10000 km. This is much smaller than the typical length-scale of flare interaction with the stellar wind thus, enabling particle confinement. 

The typical synchrotron cooling timescale of particles, $\tau \sim \frac{5.1\times10^{8}}{\gamma_{0}~B^{2}}$~s where, $\gamma_{0}$ is the Lorentz factor and $B$ is magnetic field at the interaction location. For relativistic particles (i.e. those that will emit broad-band synchrotron emission in the radio band), $\gamma_{0} \sim 10^{2}$ and for $\tau \geq$ 10$^{3}$~years, $B\leq$10$^{-4}$~G. For electrons with GeV energies, This value is consistent with the expected wind magnetic field at the location of the interaction \citep{walder2012}. Hence, the shock will emit its energy budget at a relatively constant radio luminosity over its lifetime, instead of evolving significantly on the synchrotron cooling timescale. We can use this approach to scale the emitted energy for a range of magnetic fields of the neutron star. Assuming either a $1$\% or $0.01$\% efficiency of conversion to radio emission during the interaction with the stellar wind, we can compute the luminosity of the PRS for a range of survival times of the PRS (note that the lifetime does not equal the age of the system but instead to total time the PRS is active). Figure~\ref{fig:magPRS} shows the results of such an analysis. We can see that in this case, an extremely large magnetic field is required (10$^{16}$~G) to power a PRS for a few hundred years at their observed luminosities. This again suggests that only FRB progenitors that are young will be able to power a PRS at the luminosity comparable to the PRS of FRB~20121102A. 

While the population of accelerated particles, trapped in the stellar wind's magnetic field, will also emit a higher energies, such an X-ray or $\gamma$-ray counterpart will be short lived and therefore challenging to detect. Synchrotron cooling, for instance, scales inversely with the square root of the characteristic emission frequency. Therefore, if the radio-emitting population survives for $10^3$ years, the X-ray counterpart will fade on the time scale of $\sim 10$ days. The $\gamma$-ray counterpart will be even shorter lived. As these transient counterparts are not associated with a specific radio burst, randomly catching them is unlikely, in particular in source where a PRS is already detected. Furthermore, if other processes (e.g. Inverse Compton cooling from interactions with stellar photons) speeds up the cooling, a high-energy counterpart detection is even less likely. A counterpart would similarly be expected at frequencies in between the radio and X-ray band, with an increasing lifetime towards lower frequencies.

\subsubsection{The time evolution of NMB shock models}
\label{sec:timeevo_lr}

While the previous section focuses broadly on luminosity arguments, we should also incorporate the constraints on radio variability of PRSs obtained over the past five years. For the PRS of FRB~20121102A, for instance, apparently-stochastic variability was observed~\citep{chatterjee2017}. However, no systematic decrease of its radio luminosity has been identified over a time scale of several years: it appears variable around a relatively stable specific luminosity of $2\times10^{29}$ ergs~s$^{-1}$~Hz$^{-1}$. Therefore, the NMB scenarios described above, should be able to explain such a stable luminosity with superimposed variations. Here, we will assess how the luminosity of an intra-binary shock, powered by a magnetar giant flare or by spin-down, would evolve with time. 

\begin{figure}
	\includegraphics[width=\columnwidth]{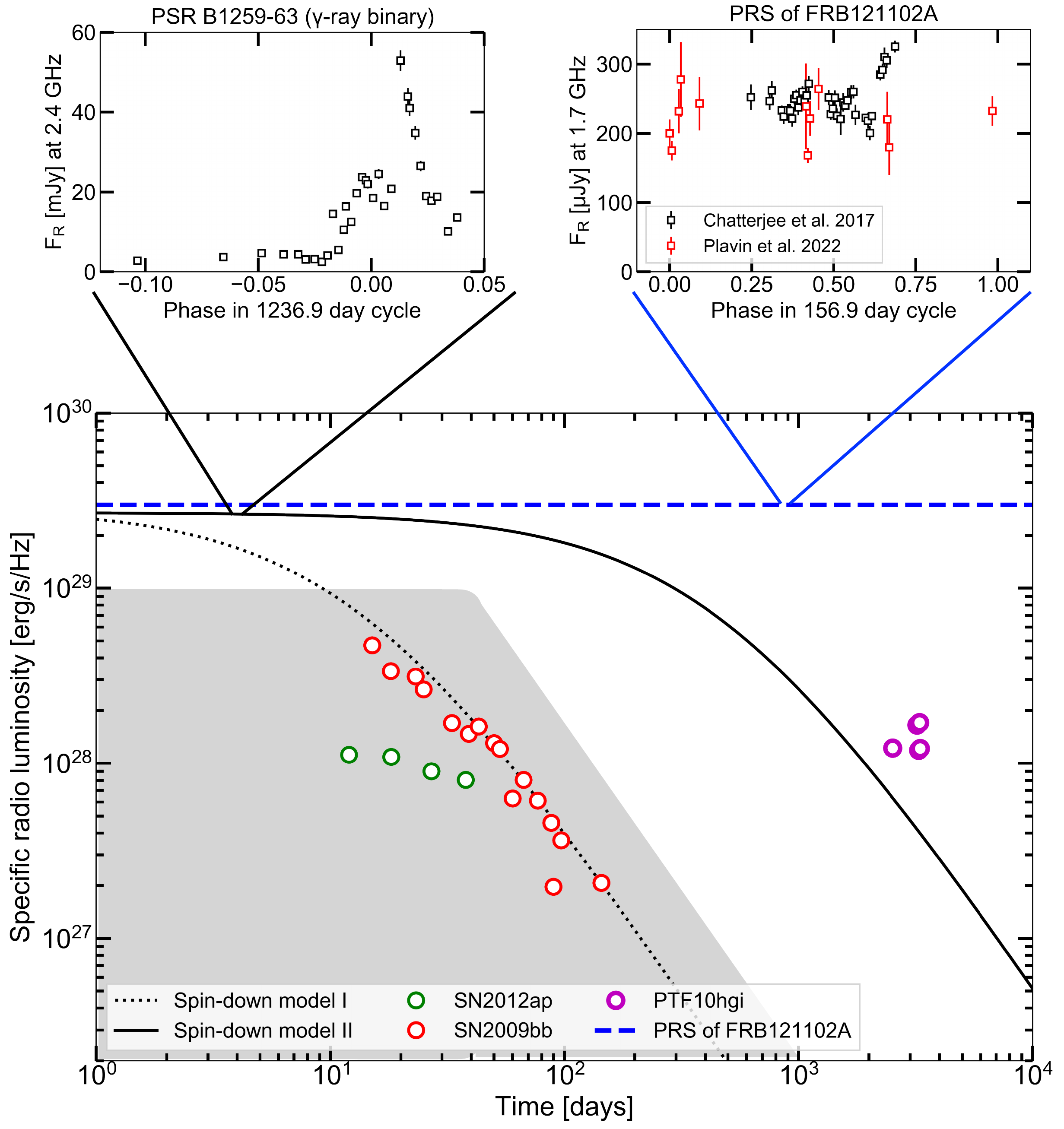}
    \caption{A schematic overview of the time evolution of the radio luminosity of several discussed scenarios for the PRS. Note that the spin-down models represent the orbitally-averaged luminosity and do not show the expected orbital modulation, as shown in the insets. The grey region indicates the typical region where type Ic supernovae are observed. See Section \ref{sec:timeevo_lr} for all details.}
    \label{fig:lr_plot}
\end{figure}

The spin-down powered intra-binary shock scenario for an NMB, as well as the supernova scenario, are depicted schematically in Figure \ref{fig:lr_plot}. The main panel of this figure shows the evolution of the radio luminosity of different sources as a function time. The circles show data from the relativistic SNe SN2008bb \citep[][red]{soderberg2010} and SN2012ap \citep[][green]{margutti2014,chakraborti2015}, as well as the late-time re-brigtening of the SLSN PTF10hgi \citep{eftekhari2019}. The light blue shaded region schematically depicts the region covered by current detections of Type Ic SNe. Finally, the blue dashed line shows the radio luminosity of the PRS of FRB~20121102A, where we plot a constant value as no systematic, long term radio evolution has been measured. As we don't expect strong time-evolution on the shown time scales for the giant-flare-powered scenario (by definition in the calculation, see above), we have not included this in the figure. 

The dotted and full black lines show the $\gamma$-ray binary type scenario, where the dipolar spin down powers the radio emission at a constant efficiency. As detailed in Appendix B, we plot the time evolution due to the decrease in spin down energy, assuming an initial magnetic field of either $3\times10^{12}$ G (case I) or $10^{14}$ G (case II) and a initial spin required to match the PRS radio (i.e. following the same scaling calculation as in the previous section). In both cases, time evolution is still expected on the time scale of several years, inconsistent with available PRS observations. This may imply that an even higher magnetic field may be required, which increases the duration of the initial luminosity plateau visible for case II; however, more likely, it indicates how the dipolar spin down is too simplistic an assumption to model the time evolution for these extreme systems; instead, therefore, magnetar flares may instead be a more favourable power source.

As stated, initial radio observing campaigns of FRB 20121102A also revealed seemingly stochastic variability around its typical radio luminosity. On the top right panel of Figure \ref{fig:lr_plot}, we show the observations by~\cite{chatterjee2017} and~\cite{plavin2022} of the PRS of FRB~20121102AA, rescaled to 1.7 GHz using a spectral index $\alpha = -0.54$ and folded on the period of its FRB activity cycle (156.9 days). In the NMB scenario, the activity cycle represents the binary orbit, on which intra-binary shocks are expected to be variable: $\gamma$-ray binaries are known to be variable along their orbit, as well as unrelated to orbital phase \citep[see e.g. the example of the radio monitoring of the $\gamma$-ray binary PSR B1259-63 around periastron plotted in the top left panel of Figure \ref{fig:lr_plot}, adapted from][]{johnston1999}. Similarly, in colliding wind binaries, where synchrotron-emitting shocks arise between two massive star winds, strong orbital variability is observed due to variation in distance between the stars. For the NMB case, particularly of a spin-down powered shock, distance variations along an eccentric orbit could similarly lead to variability, as periodically a larger and smaller fraction of pulsar wind is traversing and powering the shock. In the currently published observations of the PRS of FRB 20121102A, we do not distinguish clear variability on the activity cycle time scale. However, these observations either suffer from low signal-to-noise (red points) or cover only part of a single orbit (black points). Therefore, we propose that detailed and systematic radio monitoring across activity windows is required to fully study the unexplained PRS variability of FRB 20121102A. 

In the NMB scenario, PRS time evolution is automatically tied to spatial movement and therefore also RM and DM variability. The PRS, in both intra-binary shock scenarios, would move with the FRB source, as the entire binary travels from its birth place. In the spin-down-powered case, the long-term decrease in radio brightness would, eventually, render the PRS undetectable: indeed, FRB 20180916B may have travelled $\sim 250$ pc over $>800$ kyr to reach its current position \citep{tendulkar2021} and does not have an associated PRS at the level of FRB 20121102A and FRB 20190520B. The position of the PRS of 20121102A, on the other hand, remains within the half-light radius of its probable natal star-forming region \citep[at current positional accuracy][]{marcote2017}. This consistency implies a relatively young age ($<5.8$ Myr assuming a velocity of $34$ km s$^{-1}$), fitting in the above scenario. If the intra-binary shock is instead powered by giant magnetar flares, the PRS may be expected to remain detectable on longer time scales (tens of kyr). 

If the intra-binary shock PRS is only detectable at relatively young ages, its detection should be associated with high RM and DM and fast evolution in both quantities, as the system remains embedded in the SNR. If repeating bursts are a signature of a young FRB source, one would expect to find a detectable PRS only for repeating sources (i.e. due to a physical reason instead of an observational bias driven by the localisations of repeaters). However, inversely, a young source does not necessarily imply a detectable PRS: for both intra-binary shock scenarios, the radio luminosity depends too strongly on the natal properties of the pulsar, as well as orbital properties such as eccentricity and separation, to always render a detectable persistent counterpart. Such differences may also manifest as differences in radio variability between PRSs, either in the decrease versus time or orbitally superimposed -- therefore, FRB 20190520B may show different time evolution than FRB 20121102A does, in the NMB scenario. 

From these comparisons, we conclude that the physical size limits, specific luminosity and a lack of high-energy counterpart to FRB PRSs may be explained in a NMB scenario. In that case, the PRS would be more likely powered by giant magnetar flares than pulsar spin down, given the lack of observed decrease in PRS luminosity on time scales of years. Orbital variations of the radio luminosity may instead be expected on the time scale of the FRB activity cycle, which future observational campaigns may attempt to detect. Similarly, deep X-ray observations with future observatories may search for high-energy counterparts of the radio-emitting shocks. This scenario also predicts that PRSs are only detectable for young FRB sources. We do note the important caveat that intra-binary shocks will be highly dependent on the properties of the binary — therefore, the inferences drawn here from FRB 20121102A may not be representative of the entire population. 

\section{Discussion}
\label{sec:discussion}
While the FRB observables discussed in the previous two sections may be explained to certain extent in a NMB scenario, we briefly discuss a couple of further limitations and caveats here, while also proposing a handful of future observables that may test or rule out a NMB scenario. 

Firstly, in our considerations, we have assumed the host galaxy properties to be similar to the Milky Way. While hosts of repeating FRBs have different morphologies, SFRs, and physical sizes compared to the Milky Way, the expected NMB distribution and density in the Milky Way (based on the X-ray luminosity as a proxy for HMXBs) is consistent with the properties of host galaxies of the published repeating FRBs (see Figure~\ref{fig:galaxies}). Therefore, we believe that assuming a Milky Way like host to perform our simulations is a reasonable assumption. 

Recently,~\cite{pastor-marazuela2021} showed that for FRB~20180916B, the burst activity window is frequency dependent: bursts at lower observing frequencies appear earlier than those at high frequency, while lasting for a longer duration as well~\citep[see also]{pleunis2021}. These results suggest that if the FRBs are produced in the magnetosphere of the compact object, they will be affected by the intra-binary medium of the binary system that will completely contradict the observed chromaticity in FRB~20180916B. However, we do note that these observations do not completely rule out the binary scenario. Very recently,~\cite{li2021} have introduced a model where FRBs are generated by a neutron star in the Be/X-ray binary system, claiming that accretion induces star-quakes on the neutron surface resulting in FRBs. Their simulations have already shown that free-free absorption from the circum-stellar disk can mimic the chromatic burst activity of FRB~20180916B without invoking a non-binary model. Furthermore, several authors have also suggested binary models that can explain the low frequency activity of FRB~20180916B~\citep[see][and the references therein]{deng2022}.

The recent localisation of the repeating FRB~20200120E to a globular cluster in M81~\citep{kirsten2021} is very challenging to incorporate into a NMB scenario. While any magnetar-based FRB model will require an exotic formation channel to explain the presence of a magnetar in such an old population, the presence of massive binary companion is unlikely -- whether in a natal binary system or through dynamical capture. In any FRB mechanism that requires the presence of a binary companion, such as the ones discussed in the previous paragraph, the globular cluster localisation of FRB~20200120E implies different mechanisms across the FRB population. In that case, no PRS would be expected for FRB~20200120E through the mechanisms considered in Section \ref{sec:PRS}; in other words, if a PRS of FRB~20200120E is detected with similar properties to the other PRSs, it argues against the interpretation presented in Section \ref{sec:PRS} and a NMB model. Similarly, our discussions of the RM and DM variations (Section \ref{sec:dmrm}) requires that FRB 20200120E should not show similar variability. If the origin of FRB activity windows is related to the binary period, FRB~20200120E should also not show such periodicity in its activity. Observations so far have not revealed a periodicity in this repeating FRB~\citep{nimmo2022}.

More generally than just for FRB 20200120E, the evolution of the RM and DM in repeaters is a testable prediction for the NMB scenario: a periodic variation will be seen in the RM and DM variation that is correlated with the burst activity window. We do note that the detectable modulation in the DM and RM with orbital phase is very much dependent on the orbital inclination angle and hence may not be true for all repeating FRBs~\citep{pleunis2021}. Furthermore, if the orbits are circular, we do not expect any variation in the DM/RM due to the orbit. Similar tests can be performed with the time-dependent evolution of the PRS, which is expected to be irregular or orbital on short time scales but decaying on longer (i.e. a large number of orbits) time scales as the magnetar spins down. Finally, if a close-by repeater is localised, we may expect to detect an optical persistent counterpart in the form of the massive OB companion star. If, on the other hand, the increase in the number of localised FRBs shows that a substantial fraction of them reside in globular clusters, a NMB scenario for FRBs will become less likely. 

\section{Conclusions}
\label{sec:conclusions}

In this paper, we critically examined a scenario where a FRB progenitor is a neutron star in a binary orbit around a massive star (referred to as an NMB system). We discussed the expectations of such a scenario for different properties of repeating FRBs, such as the offsets from the birth sites of the FRB progenitors (assuming that the FRB progenitor is highly energetic neutron star), evolution of RM and DM as a function of time, or the existence of a PRS in such systems. 

Firstly, we considered the expected birth rates of HMXBs, as a proxy for NMBs, in comparison with FRBs and conclude that the discrepancy between the two can be reconciled if a very small subset of NMBs host a neutron star capable of producing FRBs. We report an upper limit to this fraction of 1$\%$ that is consistent with the current observations and expected rates of young, energetic neutron stars in our Galaxy. We also investigated any potential correlations between host galaxy properties of well localised repeaters and galaxy properties that dictate formation of NMBs. We show that current repeater hosts all fall in the region of galaxies expected to host a larger population of NMBs, although the results remain inconclusive at this point due to low number statistics. 

Using simple models, we then discussed how such a population of NMBs can mimic the offsets from star forming regions seen for a few repeaters, as well as the observed RM and DM evolution for repeating FRBs; the large diversity seen in the RM and DM variations of known repeaters may, in this scenario, be explained by the stellar wind of massive companion star. Finally, we comment on the potential for detecting a PRS in a NMB. We consider several possibilities and come to the conclusion that the PRS could be produced by intra-binary shock with the massive star's wind, powered by the neutron star's spin down or giant magnetar flares. We find that the observed stability of PRS emission over years time scales fits best with the latter power source. 

With these discussions, we have aimed to provide a framework to discuss future FRB observations in the context of NMB-type scenarios. In conjunction, we currently conclude that larger numbers of localisations and observations of repeaters will be necessary to conclusively suggest or rule out a connection between (repeating) FRBs and NMBs.

\begin{acknowledgements}
 The authors thank the anonymous reviewer whose comments significantly improved the manuscript. The authors thank Jason Hessels and Joeri van Leeuwen for helpful comments on a draft version of this work. KMR would like to thank Zorawar Wadiasingh for useful discussions. KMR acknowledges funding from the European Research Council (ERC) under the European Union's Horizon 2020 research and innovation programme (grant agreement No 694745). KMR acknowledges support from the Vici research programme ``ARGO'' with project number 639.043.815, financed by the Dutch Research Council (NWO). JvdE is supported by a Lee Hysan Junior Research Fellowship awarded by St. Hilda's College, Oxford. We gratefully acknowledge support from the Leids Kerkhoven-Bosscha Fonds (LKBF). The research leading to these results has received funding from the European Union’s Horizon 2020 Programme under the AHEAD2020 project (grant agreement n. 871158). 
\end{acknowledgements}

%
\bibliographystyle{aa} 
\bibliography{example.bib} 
%
\begin{appendix}
\section{Assumptions for the calculation of DM and RM in supernova remnant shell}

\subsection{Ionization Fraction}
We assume that supernova ejecta form a shell of uniform density with a thicker layer of swept up material on the edge of the shell. The remnant is assumed to be predominantly made up of Hydrogen. We consider four main stages of the supernova evolution: 1) free expansion phase; 2) adiabatic Phase (Sedov-Taylor); 3) radiative phase (snowplough); and 4) dispersion phase. 
For the free expansion phase, we assume that the reverse shock was created $\sim$100 years from the event and it takes about 900 years to reach the newly-formed compact object. This is assuming a forward shock velocity of 5226~km~s$^{-1}$ and a reverse shock velocity of 1700~km~s$^{-1}$~\citep{micelotta2016}. Typically, during the early stages of the free expansion stage, the recombination rate for singly ionized atoms is high ($\sim$10$^{-12}$~cm$^{-3}$~s$^{-1}$) so the rate becomes comparable to the ionization rate at a time,
\begin{equation}
\begin{split}  
    t_{\rm rec} = 4.5~\left(\frac{A_{i}}{20}\right)^{-0.5}~\left(\frac{n_{e}}{n_{i}}\right)^{0.5}~\left(\frac{\alpha}{10^{-12}~\text{cm}^{-3}~\text{s}^{-1}}\right)^{0.5}~\left(\frac{M_{\rm ej}}{1.4~M_{\odot}}\right)^{0.5} \\
    ~\left(\frac{v_{\rm ej}}{10^{4}~\text{km}~\text{s}^{-1}}\right)^{-1.5}~\text{yr},
\end{split}
\end{equation}
where $A_{i}$ is the mean atomic weight, $n_{e}$ is the electron density, $n_{i}$ is the ion density, $\alpha$ is the recombination rate, $M_{\rm ej}$ is the ejecta mass and $v_{\rm ej}$ is the ejecta velocity.
For any time $t > t_{rec}$, recombination looses its significant effects on the ionization fraction of the ejecta. However, by that time, the photon radiation field is weak enough that most of the ejecta tends to be neutral. Once the reverse shock is created, it propagates through the ejecta and the ejecta is optically thick to the photoionizing radiation of the reverse shock until a time,
\begin{equation}
\begin{split}
    t_{\rm opt} = 140~\left(\frac{A_{i}}{20}\right)^{-0.5}~\left(\frac{a_{\nu}}{10^{-12}~\text{cm}^{-3}~\text{s}^{-1}}\right)^{0.5}~\left(\frac{M_{\rm ej}}{1.4~M_{\odot}}\right)^{0.5}~\\
    \left(\frac{v_{\rm ej}}{10^{4}~\text{km}~\text{s}^{-1}}\right)^{-1}~\text{yr},
    \end{split}
\end{equation}
where, $a_{\nu}$ is the photoionization cross-section~\citep{draine2011}. For our case, using numbers for Cas~A from~\cite{micelotta2016} and assuming the $n_{e}$=$n_{i}$ for Hydrogen, we estimate $t_{\rm rec}$ and $t_{\rm opt}$ of 15 and 922 years respectively. This suggests that the ejecta is mostly neutral until the reverse shock reaches the core and completely ionizes the ejecta within the shell. A detailed treatment of the evolution of the ionization fraction is beyond the scope of this paper and hence we assume a fraction of 10$\%$ until the reverse shock reaches the core of the remnant. These equations and assumptions are adopted from~\cite{hamilton1984}.

\subsection{Velocity of the ejecta}
We assumed a different velocity evolution for the different phases of the supernova. During the free expansion phase, we assume the velocity to be completely constant at 5226~km~s$^{-1}$. This assumption holds true until the swept-up mass becomes equal to the ejecta mass, which happens around t$\sim$700-1000 years for type-II supernovae~\citep{draine2011}. During the adiabatic phase, the swept up mass is much larger than the ejecta mass, causing the ejecta to deccelerate. Sedov (1959) provides a solution for this phase such that the shock pressure is proportional to the ideal gas pressure of the monoatomic gas. Hence, the velocity of the ejecta,
\begin{equation}
    v_{\rm ej} \approx \left(\frac{2.94~E_{*}}{3\pi~\rho_0}\right)^{0.5}~R_{s}^{-1.5},
    \label{eq:v_ad}
\end{equation}
where $E_{*}$ is the total energy released in the supernova event, $\rho_0$ is the mass density of the ejecta and $R_{s}$ is the radius of the shell. We use equation~\ref{eq:v_ad} to evolve the SN-shell from $t=10^3$ to $t=2.5\times10^4$ years. For the radiative phase, we can use conservation of momentum of the shell to get the velocity of the shell at any time $t$,
\begin{equation}
    v_{\rm ej} \approx V_{0}~\left(1 + \frac{4V_{0}t}{R_{0}}\right)^{-0.75},
\end{equation}
where $V_{0}$ is the downstream velocity, $R_{0}$ is the shell radius. The velocity evolves in this way from $t=2.5\times10^4$ to $t = 10^6$ years. In the dispersion phase, we assume that the velocity of ejecta is almost equal to the velocity in the circumstellar medium (10~km~s$^{-1}$)~\citep{crawford1997}.

\subsection{Magnetic Field}

The magnetic field evolution in a supernova remnant can be parametrized as,
\begin{equation}
    B_{0} \propto \sqrt{\rho_{0}~V_{s}^{2+x}},
\end{equation}
where $V_{s}$ is the shock velocity, $\rho_{0}$ is the mass density of
the material and $x$ parameterizes the relationship between the
magnetic field and the properties of the shock (temperature,
pressure, density etc.). Typically, the density profile of the
ejecta follows $R_{s}^{s}$ where $R_{s}$ is the shock radius and
$s$=$-$2. The radius roughly evolves as $t^{m}$ where $m$=1.0.
Hence,
\begin{equation}
B(t) = B_{0} \left(\frac{t}{t_{0}}\right)^{\frac{(2-s)m}{2}+1 -\frac{x(m-1)}{2}},
\end{equation}
where $B_{0}$ is the magnetic field at time $t_{0}$. the values for $B_{0}$ and $t_{0}$ were again adopted 
from~\cite{micelotta2016} for CasA. We assume x = 1.0 and use this equation to evolve the magnetic field in the 
supernova remnant with time. The detailed mathematical derivations of the equations, as well as the values of 
different parameters presented here, are provided and discussed in~\cite{Vink2020}.

\subsection{Simulations}
For the final simulations, we split time from 0 to $5\times10^6$ years into steps of 10 years. In order to compute
the electron density, we assume the number density and the forward shock radius for Cas A from Table 1 
from~\cite{micelotta2016}. Then, under the assumption that the entire shell was ionized and had this uniform 
density, we compute the total number of electrons within the shell. We use this number them to compute the electron
density for any time step in our simulation depending on the radius of the shell at that time step. For each time
step, we compute the radius, the density, the magnetic field and the distance travelled by the NS based on its kick 
velocity. We then use those values to compute the DM and RM for that given time step as measured by the observer.

\section{Time evolution of spin-down powered model}
\label{app_lr}
To calculate the time evolution of the spin down powered PRS luminosity, we assume for simplicity a dipole model where $B = C_0\sqrt{P \dot{P}}$; when $C_0 = 3.2\times10^{19}$ G/$\sqrt{s}$, the magnetic field is given in units of G. Assuming negligible changes in magnetic field strength, the spin period will evolve as $P(t) = \sqrt{2B^2/C_0^2 + P_0^2}$, where $P_0$ is the initial spin period. The aforementioned relation between spin, spin down and magnetic field then yields the time-dependent spin-down rate; combined, we can estimate the time-evolution of the spin down energy as $-\dot{E} = 4\pi^2 I \dot{P}(t)/P(t)^3$, where $I$ is the pulsar's moment of inertia. Finally, the spin down energy is converted to radio luminosity assuming the same, constant efficiency as PSR B1259-63~\citep{johnston1999}, for which a spin down energy of $8\times10^{35}$ erg/s yields a specific radio luminosity of $3\times10^{20}$ erg/s/Hz. 

\end{appendix}

\end{document}